\documentclass[12pt]{iopart}

\usepackage{graphicx}
\begin{document}

\title{The Next Decade of Physics with PHENIX}

\author{Anne M. Sickles for the PHENIX Collaboration}

\address{Brookhaven   National Laboratory, Upton NY 11973}
\ead{anne@bnl.gov}
\begin{abstract}
The first decade of RHIC physics 
and the first heavy ion running at the LHC have produced a wealth 
of data and discoveries. It is timely to now 
evaluate what has been learned and ask what compelling new
questions have been raised. In this talk, several key unanswered
questions about the properties of the strongly coupled quark 
gluon plasma and the distribution of partons inside nucleons 
and nuclei will be discussed along with
how they can be addressed experimentally. 
The PHENIX Collaboration has developed a plan
for upgrading the experiment in order to address these new
questions.  The current status of these plans will be presented.

\end{abstract}



\section{Introduction: The First Decade of PHENIX}
The main goal of the heavy ion program at the Relativistic Heavy Ion Collider (RHIC)
is to produce and study the quark gluon plasma.  
The first decade of RHIC running have been very successful.  The collaborations
have established that a new form of matter was created~\cite{wp}.
The study of that matter, the strongly interacting quark gluon plasma
(sQGP), has led to an emerging picture of matter which
behaves hydrodynamically and is nearly opaque to hard partons.

As a community, we are now working toward building a coherent understanding 
of the sQGP.  The RHIC collaborations were asked to develop plans
for the next ten years of RHIC operations.
These proceedings report on the current status of the PHENIX Decadal Planning
process.  The complete Decadal Plan, submitted in October 2010, is available 
here~\cite{decadal_plan}.

\section{New Questions in Heavy Ions}

As a vital part of planning for the future the PHENIX Collaboration discussed what
compelling questions remained after the first decade of RHIC running and what
new questions have been raised by our current knowledge.  Within the realm of 
heavy ion physics five compelling questions were identified:
\begin{itemize}
\item are quarks strongly coupled to the QGP at all distance scales?
\item what are the detailed mechanisms for parton-QGP interactions and responses?
\item are there quasiparticles at any scale?
\item is there a relevant screening length in the QGP?
\item how is rapid equilibration achieved?
\end{itemize}
For the first two questions, jets, including heavy flavor tagged jets
are an essential tool to probe the matter.  The screening length
is addressed by high quality quarkonia measurements.
One observable relevant to the  final question
is the $v_2$ of direct photons.  In the time since the Decadal
Plan was written PHENIX has submitted such a measurement for
publication~\cite{ppg126}.  Additional observables sensitive to
equilibration are under discussion.

\subsection{Hard Probes at RHIC in the Age of the LHC}
In November 2010 the Large Hadron Collider (LHC) took Pb+Pb
data for the first time at $\sqrt{s_{NN}}$=2.76 TeV.  
Many results from the ALICE, ATLAS and CMS Collaborations were shown at this 
conference.  Given the lower collision energy at RHIC, it is natural to 
ask what role hard probes, especially jet measurements,
play at RHIC.  

The question of the mechanisms by which hard probes interact with the sQGP
is open despite very detailed measurements from the first decade of RHIC data.
With the large difference in energy density at the two machines 
RHIC and LHC together  can help resolve this puzzle more effectively than
either can alone.   As the LHC experiments repeat measurements that had been
previously done at RHIC (such as single particle
$R_{AA}$ measurements), new insights are gained.  Similarly when
high quality jet measurements in heavy ion collisions are done at RHIC,
the insights gained from LHC jet results will be tested and refined.
Here jet measurements refers to  a broad class of measurements that include
not only single jet suppression factors, but also reconstructed
dijet correlations, heavy quark jets and $\gamma$-jet correlations.  A coherent
understanding of parton-matter interactions will be able to 
describe these observables across a wide range of energy densities.

\section{sPHENIX Detector Concept}
The central driver for the sPHENIX mid-rapidity detector design is jet measurements.
The sPHENIX detector philosophy rests on three points: large rate, large acceptance and
large calorimetric coverage.  While this philosophy is
central to jet measurements in high energy physics experiments, 
this combination does not currently exist at RHIC.

\begin{figure}
\centering
\includegraphics[width=\linewidth]{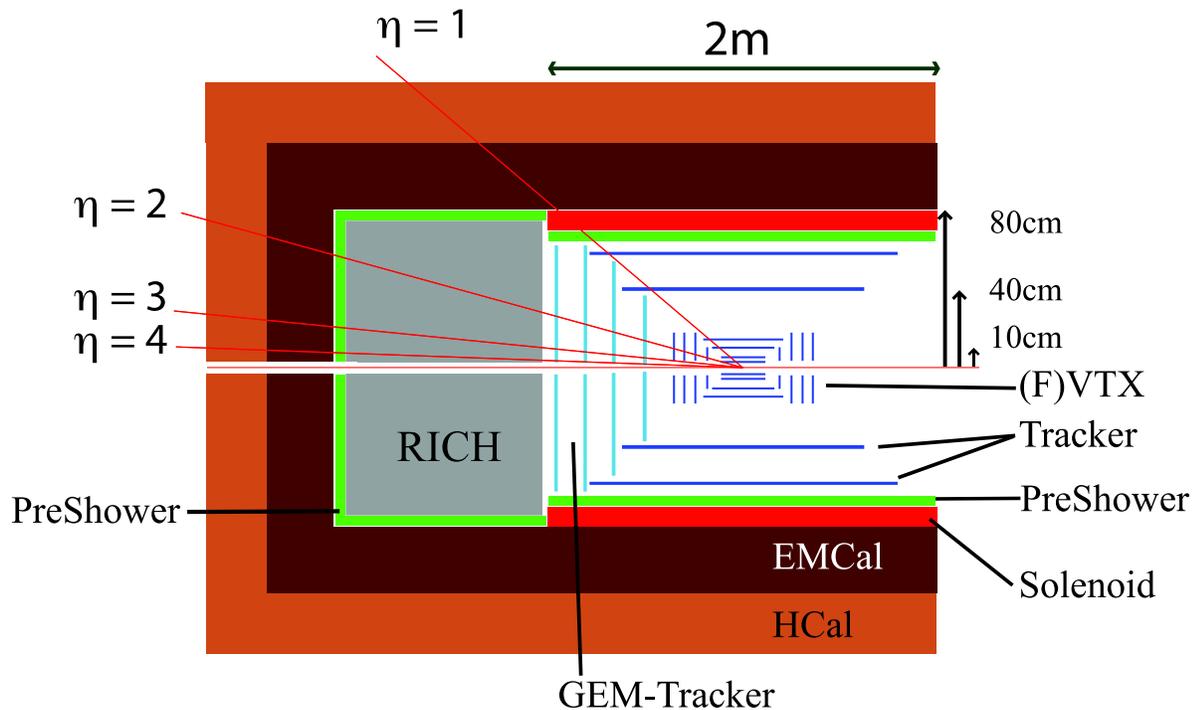}
\caption{A strawman sketch of the sPHENIX detector.  Some
features of the detector are discussed in the text.}
\label{sphenix}
\end{figure}

A cartoon drawing of a strawman sPHENIX design is shown in Fig.~\ref{sphenix}.
The existing Barrel Silicon Vertex detector (installed for the 2011 RHIC
Run) and the Forward Vertex Detector (to be installed in Summer 2011)
are retained from the present PHENIX design.  Additional silicon
tracking at a radius of $\approx$40cm is added to improve the momentum
resolution and track finding capabilities.  The central magnet is
currently envisioned to be a 2T solenoid.  Simulations performed suggest
that this should be sufficient to separate the $\Upsilon_{2S}$ and $\Upsilon_{3S}$
states from the $\Upsilon_{1S}$.  The electromagnetic calorimeter concept involves
a preshower as part of the
electron identification and $\pi^0$ reconstruction~\cite{decadal_plan}.
While the calorimeter
would be very compact, a small Moli\`ere radius ($R_M\approx$ 2cm) 
design leads to reasonable occupancies in
central Au+Au events.  
The hadronic calorimetry is a central part of the jet measurements, providing
the best possible measurement of the jet energy and the fluctuations in the 
underlying event and prevents fake high momentum tracks from seeding
fake jet signals.

Specific technology choices had to be used in the strawman design in order
to do realistic simulations, however the actual technology choices are still
under active discussion.  The design of the forward rapidity upgrades is discussed 
further in Ref.~\cite{decadal_plan}.

The sPHENIX goal is to be able to take nearly the entire delivered luminosity for
hard probes in heavy ion collisions.  
Based on luminosity projections we anticipate something on the order
of 50B events sampled in a Au+Au run.  Notably, $\approx$ 25B
minimum bias events could be recorded.  This vast data sample provides rates to do not only light
flavor jet measurements, but also di-jet measurements, heavy quark tagged jets,
$\gamma$-jets and quarkonia.  

\begin{figure}
\begin{minipage}{0.52\linewidth}
\includegraphics[width=\linewidth]{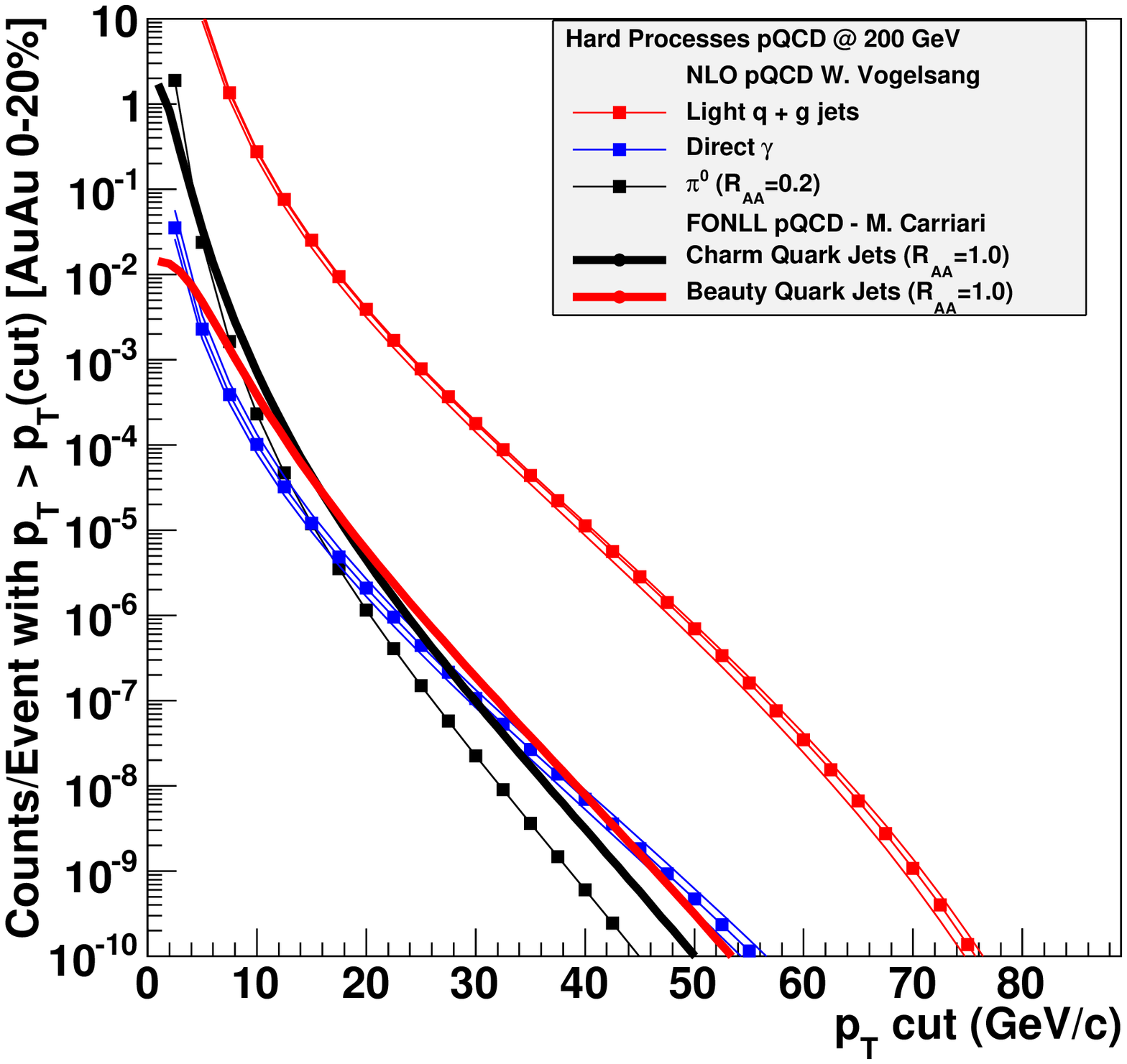}
\end{minipage}
\begin{minipage}{0.48\linewidth}
\includegraphics[width=\linewidth]{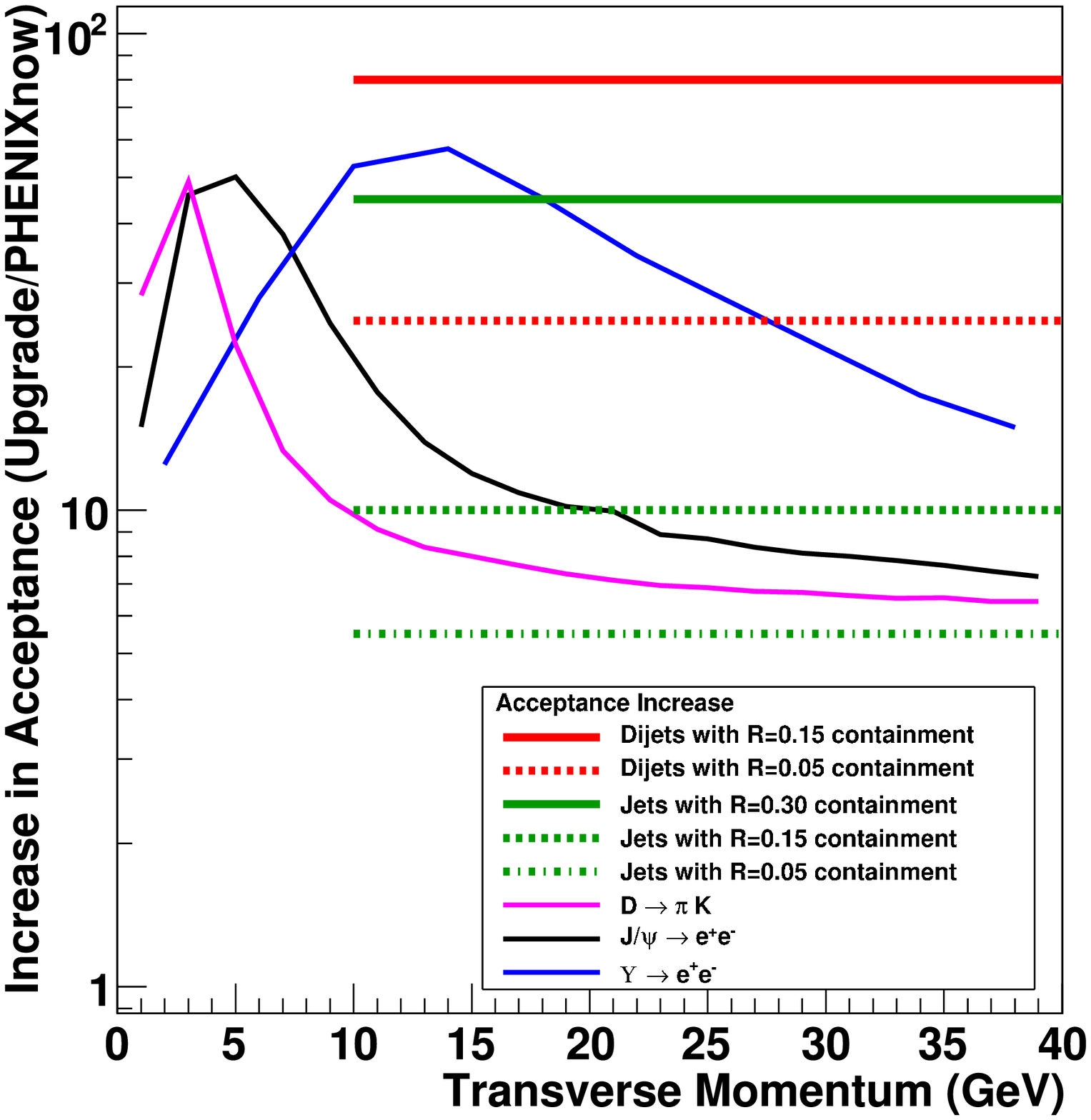}
\end{minipage}
\caption{(left)Calculated rates for hard probes production in central Au+Au collisions.
Figures is adapted from Ref.~\cite{decadal_plan}.}
\label{rates}
\end{figure}

Figure~\ref{rates}(left) shows the rates for jets, $\pi^0$s, direct photons
and heavy quarks expected in central Au+Au events.  Plotted on the y-axis
is the number of counts expected per event for $p_T \ge p_T(cut)$
where $p_T(cut)$ is the value on the x-axis.
Specifically, if one wants on order of 1000
counts for the various channels this allows light quark jets up to about
60GeV/c and $\approx$30GeV/c for charm and bottom jets (provided
they are able to be tagged).
Fig~\ref{rates}(right) shows the acceptance increase from
the current PHENIX acceptance to the sPHENIX acceptance given
a pseudo-rapidity acceptance of $|\eta|<$1.  
Acceptance increases by factors of 10 to nearly 100 are seen for various probes.

\section{Conclusions}

The sPHENIX plan provides a detector that is well matched to compelling
heavy ion physics goals.  Additionally, RHIC is a collider known for its
versatility.  This detector is well matched to polarized p+p and d+Au collisions
to study the spin structure of the nucleon and the cold nuclear matter effects
as well.  Future goals in these programs especially constrain the forward capabilities 
of this detector and are discussed in detail in the Decadal Plan~\cite{decadal_plan}.
Additionally, as part of an electron ion collider this detector could be
used for some e+A and e+p measurements as well.  

Precision jet measurements in heavy ion collisions at RHIC will be complementary
to those taken at the LHC.  
Having two machines at well separated collision
energy, both with the rates, acceptance
and detectors capable to study
hard probes such as jets and quarkonia provides excellent opportunities to gain
insight
into the puzzles that now exist, particularly energy loss and
screening lengths and be ready to confront new surprises.

\section{References}
\bibliographystyle{iopart-num}
\bibliography{sickles_qm11_proceedings}
\end{document}